\documentstyle[aps,pra,amsmath,graphicx,multicol]{revtex}
\newcommand{\bra}[1]{\left\langle#1\right|}
\newcommand{\ket}[1]{\left|#1\right\rangle}
\newcommand{\scp}[2]{\langle #1\!\mid\!#2\rangle}
\newcommand{\ave}[1]{\langle#1\rangle}
\newcommand{\fs}{{\tilde f}}
\newcommand{\ofs}{(1+\tilde f)}
\newcommand{\ms}{{\tilde m}}
\newcommand{\f}{f^{\vphantom\ast}}
\newcommand{\Gs}{\tilde G}
\newcommand{\gs}{\tilde g}
\newcommand{\as}{\hat{\tilde a}}
\newcommand{\adag}{\hat{a}^\dag}
\newcommand{\al}{\alpha^{\vphantom\ast}}
\newcommand{\bx}{{\mathbf{x}}^{\vphantom\dag}}
\newcommand{\xx}{{\mathbf{x}}}
\newcommand{\yy}{{\mathbf{y}}}
\newcommand{\fsp}{\tilde f^{\vphantom\ast}}
\newcommand{\msp}{\tilde m^{\vphantom\ast}}
\newcommand{\gls}{{\stackrel<>}}
\newcommand{\lgs}{{\stackrel><}}
\newcommand{\av}{\hat{a}^{\vphantom\dag}}
\newcommand{\hc}{\mathrm{H.c.}}
\newcommand{\ant}{~and~}
\newcommand{\bigave}[1]{\big\langle#1\big\rangle}
\newcommand{\der}[1]{\frac{d}{d#1}}
\newcommand{\tr}{{\mathrm{Tr}}}
\newcommand{\qandq}{\quad\text{and}\quad}
\newcommand{\ep}{\epsilon^{\vphantom\dag}}
\newcommand{\V}[1]{V_{\mathrm{#1}}}
\newcommand{\Hzero}{\hat{H}^{(0)}}
\newcommand{\HF}{_{\mathrm{HF}}}

\begin{document}

  \title{Equivalence of kinetic theories of Bose-Einstein condensation}
  \author{J.~Wachter, R.~Walser, J.~Cooper, and M.~Holland}
  \address{JILA, National Institute of Standards and Technology and
	University of Colorado, \\Boulder, Colorado 80309-0440}
  \date{Sep 24, 2001}
  \maketitle
  \begin{abstract}
We discuss the equivalence of two nonequilibrium kinetic theories that
describe the evolution of a dilute, Bose-Einstein condensed atomic gas
in a harmonic trap.  The second-order kinetic equations of Walser {\it
et al.} [Phys.\ Rev.\ A {\bf63}, 013607 (2001)] reduce to the
Gross-Pitaevskii equation and the quantum Boltzmann equation in the
low- and high-temperature limits, respectively.  These kinetic
equations thus describe the system in equilibrium (finite temperature)
as well as in nonequilibrium (real time).  We have found this theory
to be equivalent to the nonequilibrium Green's function approach
originally proposed by Kadanoff and Baym and more recently applied to
inhomogeneous trapped systems by Imamovi\'c-Tomasovi\'c and Griffin
[in {\it Progress in Nonequilibrium Green's Functions,} edited by
M. Bonitz (World Scientific, Singapore, 2000), p.\ 404].
  \end{abstract}

\pacs{PACS numbers: 03.75.Fi, 05.70.Ln}

\ifpreprintsty\else\begin{multicols}{2}\fi
\section{Introduction}
Binary collisions are the essential mechanism for the formation of a
Bose-Einstein condensate in an atomic gas.  
Moreover, many aspects of the system's dynamics require two-particle
collisions, for example, sound propagation, the damping of elementary
excitations, and the very mechanism that leads to the quantum phase
transition---evaporative cooling.
However, the conventional
Hartree-Fock-Bogoliubov approach to generalize the Gross-Pitaevskii
equation for dilute, trapped gases includes binary collisional
interactions only as first-order energy shifts. Second-order kinetic
theories that include collisional redistributions of excited atoms
offer a more complete microscopic description of the gaseous system.

Why is a simplified kinetic description possible, when the evolution
of the Bose-Einstein condensate might involve correlations between as
many particles as the system contains?  Would not binary collisions
eventually entangle the quantum state of each atom in the system with
that of every other atom?  Fortunately, such complexity is not
necessary to describe the measurable properties of a dilute, weakly
interacting gas, because the duration of a collision, $\tau_0$, is
very short compared to the essentially interaction-free oscillation in
the external potential between isolated collision events~\cite{Zubarev1}.

Because of this characteristic separation of time scales, correlations
that arise during an individual collision decay rapidly before the
next collision takes place.  This rapid decay, in turn, implies the
possibility of a Markov approximation, which assumes that only the
current configuration of the system determines its future evolution.
Furthermore, this decay of correlations allows us to parametrize the
system's state by a reduced set of master variables, because we are
interested in the system's time evolution only on the kinetic time
scale, i.\thinspace e., for times large compared to the duration of a
collision~$\tau_0$. This reduced description with a set of master
variables is possible, because for kinetic times the higher-order
correlation functions can be expressed as functionals of these
variables~\cite{Akhiezer}.

This set of master variables is common to both kinetic theories we
will discuss: In the Kadanoff-Baym approach, abstract real-time
Green's functions parametrize the condensating gas, whereas in the
Walser {\it et al.}\ case~\cite{Walser2001a}, single-time density
matrices, which contain the physical density and coherences of thermal
atoms, as well as the mean field, represent the system. The
equivalence of these two approaches is a general principle in
nonequilibrium statistical
mechanics~\cite{Morozov1999a,Zubarev2}. However, it is not trivial to
verify this fact in detail by explicitly connecting the complementary
microscopic equations. Strictly speaking, we state equivalence after
the Kadanoff-Baym theory has been restricted to single-time quantities
using the Markov approximation.

We present the formulation of the quantum kinetic theory of dilute
Bose-Einstein condensed gases in terms of nonequilibrium, real-time
Green's functions and their Kadanoff-Baym equations of
motion~\cite{Kadanoff}, which were generalized in
Refs.~\cite{Kane1965a,Hohenberg1965a} to include the condensate.

By transforming these equations to the single-particle energy basis
and taking the single-time limit of the two-time Green's functions by
means of the Markov approximation, we reproduce the equations of
motion of the Walser {\it et al.}\ kinetic theory as presented in
Ref.~\cite{Walser2001a}, thus providing an independent confirmation of
these equations.  Following Imamovi\'c-Tomasovi\'c and
Griffin~\cite{Imamovic2000a}, we use the gapless Beliaev
approximation for the self-energies in the Kadanoff-Baym equations,
and thus prove the Walser {\it et al.}\ kinetic theory to be gapless as
well.
\section{Nonequilibrium Green's Functions}
We begin the introduction to the Kadanoff-Baym description of the
dilute Bose gas by defining its variables. Neglecting three-body
interactions, the second-quantized
many-body Hamiltonian~$\hat H$ describing the atoms is
\begin{eqnarray}
  \hat H=&&\int d\xx\int d\yy\;\adag(\xx)\;\bra{\xx}\Hzero
	\ket{\yy}\;\hat{a}(\yy)\nonumber\\
  &&+\frac12\int d\xx \int d\yy\;\adag(\xx)\adag(\yy)\;\V{bin}(\xx-\yy)\;
	\hat{a}(\yy)\hat{a}(\xx),
\end{eqnarray}
where $\adag(\xx)$ is the bosonic creation operator and $\V{bin}(\xx-\yy)$
the binary interaction potential. The single-particle Hamiltonian
\begin{equation}
  \Hzero=\frac{\hat{\mathbf{p}}^2}{2m}+\V{ext}(\hat\xx)\label{Hzero}
\end{equation}
contains the kinetic energy of a boson with mass~$m$ and the external
potential~$\V{ext}(\xx)$.

To represent the master variables in terms of nonequilibrium
Green's functions, we first write the system's degrees of freedom
in terms of spinor operators~\cite{Nambu1960a}
\begin{eqnarray}
  \hat{A}(1)=\left(\begin{array}{c}\hat{a}(1)\\\adag(1)\end{array}\right)
  \quad\text{and}\quad
  \hat{A}^\dag(1)=\left(\begin{array}{cc}\adag(1)&\hat{a}(1)\end{array}\right),
\end{eqnarray}
where we now follow Kadanoff-Baym and abbreviate $(1)\equiv(\bx_1,t_1)$. 
The master variables are then contained in the following two-time propagators:
\begin{eqnarray}
  h(1,2)&\equiv&-i\ave{\hat{A}(1)}\ave{\hat{A}^\dag(2)},\label{defh}\\
  g(1,2)&\equiv&-i\bigave{T\big\{\hat{A}(1)\hat{A}^\dag(2)\big\}},
\end{eqnarray}
where~$\ave\cdot$ denotes the grand-canonical average and~$T\{\cdot\}$ the
time ordering operator, which sorts its arguments in 
order of decreasing time. These two propagators are defined for real
times by analytic continuation of the finite-temperature propagators
for imaginary time, following~\cite[Chap.~8]{Kadanoff}.
We subtract the condensate propagator~$h$ from the full
propagator~$g$ and thus define the Green's function for the fluctuations
\begin{equation}
  \gs(1,2)\equiv g(1,2)-h(1,2).
\end{equation}
The two time orderings of~$\gs$,
\begin{equation}
  \gs^<(1,2)\equiv\gs(1,2)\quad\text{for}
	\quad t_1<t_2 \label{defgsl}
\end{equation}
and
\begin{equation}
  \gs^>(1,2)\equiv\gs(1,2)\quad\text{for}
	\quad t_1>t_2, \label{defgsg}
\end{equation}
define the generalized two-time fluctuation-density matrices. This can be
seen by explicitly writing these two time orderings in terms of the
fluctuating part~$\as(1)$ of the field operators,
\begin{equation}
  \as(1)\equiv \hat{a}(1)-\ave{\hat{a}(1)}\equiv \hat{a}(1)-\alpha(1),
	\label{defalpha}
\end{equation}
as follows:
\begin{eqnarray}
  \gs^<(1,2)&=&\left(\begin{array}{cc}\fsp_{12}&\msp_{12}\\
	\ms^\ast_{12}&\ofs^\ast_{12}\end{array}\right),
	\label{fmsdef}\\
  \gs^>(1,2)&=&\sigma_z+\gs^<(1,2),\label{fmgdef}
\end{eqnarray}
where we defined the two-time
normal~($\fs$) and anomalous~($\ms$)
averages of the fluctuations in the position basis as
\begin{equation}
  \fsp_{12}=\bigave{\as^\dag(2)\as(1)}\qandq\msp_{12}=\bigave{\as(2)\as(1)}.
	\label{fsms}
\end{equation}
In the case~$t_1=t_2$, the propagators in
Eqs.\ \eqref{fmsdef}~and~\eqref{fmgdef} correspond to the dynamical
quantities in the kinetic equations for the fluctuations given in
Eqs.\ (24)~and~(25) of Ref.~\cite{Walser2001a};
for~$t_1=t_2$, the averages in Eq.~\eqref{fsms} correspond to the
density of thermal atoms around the condensate and correlations
between these atoms. We have thus represented the condensate ($h$) and
its fluctuations ($\gs^<$) and can now look for their corresponding
evolution equations.

\section{Kadanoff-Baym Equations}
The equations of motion for the nonequilibrium Green's functions
$h$\ant$\gs^<$ are the Kadanoff-Baym equations; these equations are
equivalent to the Dyson equation.
In the second part of this section, we discuss the second-order
Beliaev approximation for the self-energies that we use.
For the condensed part of the atom cloud,
which is parametrized by the propagator~$h(1,2)$ defined in
Eq.~\eqref{defh}, we can write the Kadanoff-Baym equations
as~\cite{Kane1965a} 
\begin{eqnarray}
  &&\int_{-\infty}^\infty d\bar1 \left\{g_0^{-1}(1,\bar1)-S\HF
    (1,\bar1)\right\}h(\bar1,2) \nonumber\\
    &&=\int_{-\infty}^{t_1}d\bar1\;\left\{S^>(1,\bar1)-S^<(1,\bar1)
    \right\}h(\bar1,2)\quad\label{KKM1}
\end{eqnarray}
and
\begin{eqnarray}
  &&\int_{-\infty}^\infty d\bar1\; 
    h(1,\bar1)\left\{g_0^{-1}(\bar1,2)-S\HF (\bar1,2)\right\}\nonumber\\
    &&= -\int_{-\infty}^{t_2}d\bar1\; h(1,\bar1) \left\{
    S^>(\bar1,2)-S^<(\bar1,2)\right\}.\quad \label{KKM2}
\end{eqnarray}
We write the corresponding equations for the fluctuations
$\gs^<(1,2)$\ant$\gs^>(1,2)$ [Eqs.\ \eqref{fmsdef}\ant\eqref{fmgdef}]
around the condensate mean field as
\begin{eqnarray}
  &&\int_{-\infty}^\infty d\bar1 \left\{g_0^{-1}(1,\bar1)-\Sigma\HF 
    (1,\bar1)\right\}
  \gs^\lgs(\bar1,2)  \nonumber\\ 
  &&=\int_{-\infty}^{t_1}d\bar1\;\left\{\Sigma^>(1,\bar1)-\Sigma^<(1,\bar1)
    \right\}\gs^\lgs(\bar1,2)\nonumber\\
    &&\quad-\int_{-\infty}^{t_2}d\bar1\; \Sigma^\lgs(1,\bar1)
    \left\{\gs^>(\bar1,2)-\gs^<(\bar1,2)\right\}\label{KKF1}
\end{eqnarray}
and
\begin{eqnarray}
  &&\int_{-\infty}^\infty d\bar1\; 
  \gs^\lgs(1,\bar1)\left\{g_0^{-1}(\bar1,2)-\Sigma\HF (\bar1,2)\right\}
     \nonumber\\
  &&=\int_{-\infty}^{t_1}d\bar1\; \left\{
    \gs^>(1,\bar1)-\gs^<(1,\bar1)\right\}\Sigma^\lgs(\bar1,2)\nonumber\\
  &&\quad-\int_{-\infty}^{t_2}d\bar1\; \gs^\lgs(1,\bar1) \left\{
    \Sigma^>(\bar1,2)-\Sigma^<(\bar1,2)\right\}.\label{KKF2}
\end{eqnarray}
In Eqs.\ \eqref{KKM1}~through~\eqref{KKF2}, we use the definition of
the matrix inverse of the interaction-free propagator~$g_0$,
\begin{equation}
  g_0^{-1}(1,2)=\Big\{i\sigma^z\der{t_1}+\frac{\nabla_1^2}{2m}
  -\V{ext}(1)+\mu\Big\}\delta(1,2),	\label{defg0inv}
\end{equation}
with the third Pauli
matrix~$\sigma^z=\mathrm{diag}(\openone,-\openone)$ and an energy
shift~$\mu$, which removes mean-field oscillations.  We define the
$\delta$~function
by~$\delta(1,2)\equiv\delta(\bx_1-\bx_2)\delta(t_1-t_2)$ and
integration~$d\bar1$ as integration~$dt_{\bar1}$ over time within the
given time limits and~$d\bx_{\bar1}$ over all space. The
approximations we choose for the Hartree-Fock self-energies for the
condensate $S\HF $ and for the fluctuations $\Sigma\HF $ as well as
the second-order collisional self-energies $S^<$ and $\Sigma^<$ will
be discussed below.

Kadanoff and Baym derived these equations without including the
condensate~\cite{Kadanoff} and de Dominicis and Martin formulated a
very general mathematical
account~\cite{Dominicis1964a,Dominicis1964b}. The Green's function
formalism traces back to Schwinger~\cite{Schwinger1951a} and
originally made use of the correspondence between the partition
function and the time evolution operator in imaginary time ($e^{\beta
H}=e^{iHt}$ for $t=-i\beta$).  To get information about measurable
quantities, the dynamic variables and equations of motion were
extended to real times by analytic continuation (see
\cite[Chap.~8]{Kadanoff}~and~\cite{Baym1961a,Craig1968a,Zubarev2} for
more details).

This nonequilibrium Green's function description was developed
$40$~years ago to eventually explain the behavior of superfluid
helium~\cite{Martin1961a}. Since this description involves a
weak-coupling approximation but helium atoms are strongly interacting,
the results at that time were disappointing and, for example, could
not explain all predictions of the phenomenological Landau
model. However, since the Green's function description holds for a
dilute, weakly interacting gas, its application to Bose-Einstein
condensation in this system is more appropriate.

To complete our exposition of the Kadanoff-Baym equations
\eqref{KKM1}~through~\eqref{KKF2}, we have to choose the Hartree-Fock
and collisional self-energies.
We draw the Hartree-Fock self-energy diagrams for both the
condensate~$h$ and the thermal cloud~$\gs^<$ in Fig.~\ref{fig_HF} and write
them, respectively, as
\begin{eqnarray}
  S\HF (1,2)&=&\frac{i}{2}\int d\bar2\;v(1,\bar2)
    \tr\left\{g(\bar2,\bar2)\right\} \delta(1,2)\nonumber\\
  &&+iv(1,2)\gs(1,2)\label{defshf}
\end{eqnarray}
and
\begin{eqnarray}
  \Sigma\HF (1,2)&=&\frac{i}{2}\int d\bar2\;v(1,\bar2)
    \tr\left\{g(\bar2,\bar2)\right\} \delta(1,2)\nonumber\\
  &&+iv(1,2)g(1,2),	\label{defsigmahf}
\end{eqnarray}
with the local-time, binary interaction
potential~$v(1,2)=\V{bin}(\bx_1-\bx_2)\;\delta(t_1-t_2)$
and the matrix trace $\tr$.
When we evaluate the time-ordered propagator~$g$ at equal times, we
follow the convention $T\{\hat{a}(1)\adag(2)\}=\adag(2)\hat{a}(1)$.

\begin{figure}[t]
  \centering\includegraphics[scale=.9]{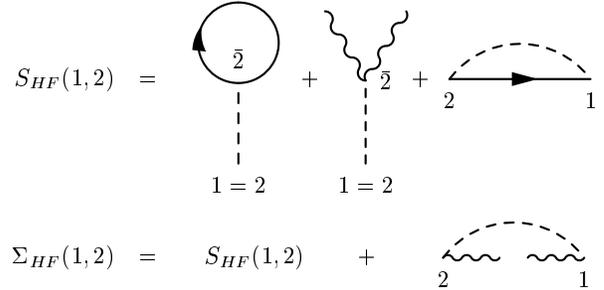}
  \caption{The first-order Hartree-Fock self-energy diagrams. The
  solid lines depict the noncondensate propagator~$\gs$, the wiggly
  lines the condensate propagator~$h$, and the dashed lines the
  interaction potential~$v$. The first two terms give the energy
  shifts due to both the mean field~$U_{f^c}$ and the normal
  fluctuations~$U_{\fs}$. The third term in~$S\HF $ gives rise to a
  factor of 2 for~$U_{\fs}$ and to~$V_{\ms}$. The fourth term which
  only appears in~$\Sigma\HF $ causes the difference in the
  mean-field shifts that are experienced by the condensate and the
  fluctuations, respectively.}
  \label{fig_HF}
\end{figure}

For the second-order collisional self-energies~$\Sigma^\lgs$ we choose the
gapless and conserving Beliaev
approximation~\cite{Imamovic2000a,Imamovic2000b,Beliaev1958a,Beliaev1958b}.
This means that, compared to Kane and Kadanoff~\cite{Kane1965a}, we
include the exchange terms, which they deliberately excluded to obtain the
simplest conserving approximation as proven in~\cite{Baym1961b}, and compared
to Hohenberg and Martin~\cite{Hohenberg1965a}, we include the terms
containing no condensate contributions, which will give rise to the
quantum Boltzmann terms for the fluctuations. 
\ifpreprintsty\else
  \end{multicols}
  \noindent
  \vspace{1pc}
\fi

\begin{figure}[f]
  \centering\includegraphics[scale=.9]{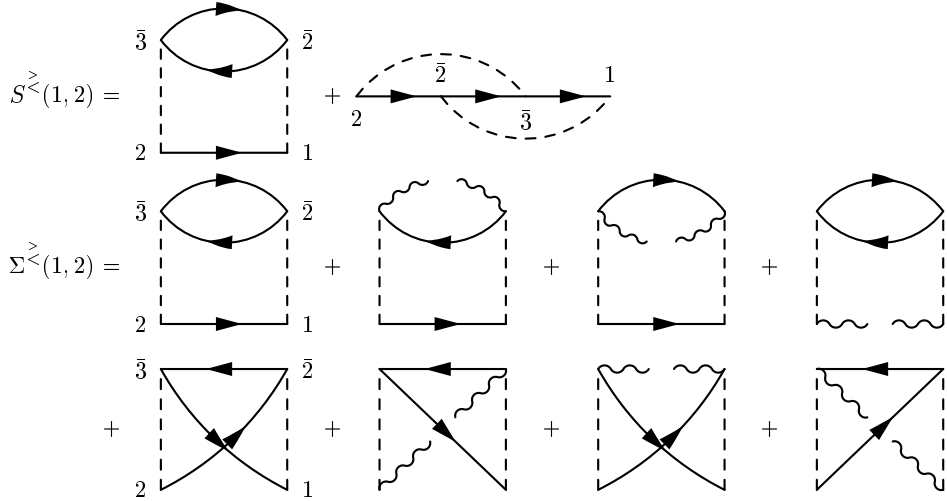}
  \caption{The second-order collisional self-energies in the gapless
  Beliaev approximation. The solid lines depict the noncondensate
  propagator~$\gs$, the wiggly lines the condensate propagator~$h$,
  and the dashed lines the binary interaction potential~$v$. The
  second diagram of~$S$ corresponds to the last four of~$\Sigma$, when
  we replace each of the three fluctuation propagators by an open
  condensate one.}
  \label{fig_coll}
\end{figure}

We depict the resulting self-energy diagrams in
Fig.~\ref{fig_coll} and represent them mathematically as
\begin{eqnarray}
  S^\lgs(1,2)=-\frac{1}{2}\int d\bar2 \int d\bar3\;
    v(1,\bar2)v(2,\bar3)&&\!\bigg[ \gs^\lgs(1,2) \tr \Big\{
    \gs^\gls(\bar3,\bar2) \gs^\lgs(\bar2,\bar3)\Big\}\nonumber\\
    &&\quad+2\gs^\lgs(1,\bar3)\gs^\gls(\bar3,\bar2)\gs^\lgs(\bar2,2)
    \bigg]\label{defscoll}
\end{eqnarray}
for the condensed part and
\begin{eqnarray}
  \Sigma^\lgs(1,2)&=& -\frac{1}{2}\int d\bar2 \int d\bar3\;
    v(1,\bar2)v(2,\bar3)\nonumber\\ 
  &&\times\bigg[\gs^\lgs(1,2) \tr \Big\{
    g^\gls(\bar3,\bar2) g^\lgs(\bar2,\bar3)
    -h(\bar3,\bar2) h(\bar2,\bar3)\Big\}
    +h(1,2)\tr\Big\{ \gs^\gls
    (\bar3,\bar2)\gs^\lgs(\bar2,\bar3)\Big\}\nonumber\\
  &&\quad+2\gs^\lgs(1,\bar3) \Big\{
    g^\gls(\bar3,\bar2) g^\lgs(\bar2,2)
    -h(\bar3,\bar2) h(\bar2,2)\Big\}
    +2h(1,\bar3)\Big\{ \gs^\gls(\bar3,\bar2)
    \gs^\lgs(\bar2,2) 
    \Big\}\bigg]\label{defsigmacoll}
\end{eqnarray}
for the fluctuations.

Instead of using lines for the matrix-valued propagators $\gs$~and~$h$
as in Fig.~\ref{fig_coll}, one can also draw diagrams for the four
elements of the matrix separately. The resulting diagrams for the
first-order and second-order Beliaev terms can be seen in
Figs. $15$~and~$17$ of Ref.~\cite{Shi1998a}, where the interaction
potential is replaced by a two-body $T$ matrix.
\ifpreprintsty\else\begin{multicols}{2}\fi

\section{Transformation to the Energy Basis}
We now demonstrate the key steps that connect the kinetic theory
presented in the previous section to the work of Walser {\it et al.}\
presented in~\cite{Walser2001a}: We rewrite the Kadanoff-Baym Eqs.\
\eqref{KKM1}~through~\eqref{KKF2} in the single-particle energy (SPE)
basis and obtain the equations of motion for the master
variables---the measurable quantities in our reduced description of
the system---in this basis, exactly as given in the Walser {\it et
al.}\ paper.

First, we define our master variables in the SPE~basis
$\{\ket{1'}\}_{1'}\equiv\{|\ep_{1'}\rangle\}_{\epsilon_{1'}}$ and
determine their relation to their position basis counterparts, the
Green's functions given in Eqs.~\eqref{defh},
\eqref{fmsdef},\ant\eqref{fmgdef}.  The time-dependent, two-component
mean-field state vector
\begin{equation}
\chi=\left(\begin{array}{c}\alpha\\\alpha^\ast\end{array}\right)
\end{equation}
is defined in terms of
$\alpha\equiv\al_{1'}\ket{1'}\equiv\sum_{1'}\ave{\hat{a}_{1'}}\ket{1'}$
and also contains the time-reversed mean field~$\alpha^\ast$.  The
time-dependent annihilation and creation operators
$\hat{a}$\ant$\adag$ transform as 
\begin{eqnarray}
  \hat{a}(1)=\scp{1}{1'}\av_{1'}\qandq
	\adag(1)=\scp{1'}{1}\adag_{1'}, \label{trfa}
\end{eqnarray}
where $\ket{1}\equiv|\bx_1\rangle$ are the position eigenstates.
The fluctuating part of the master variables is contained in the
single-time fluctuation-density matrix $\Gs^<$, which we define as
\begin{equation}
  \Gs^<=\left(\begin{array}{cc}\fs&\ms\\\ms^\ast&\ofs^\ast
	\end{array}\right),\label{defGs}
\end{equation}
with the normal fluctuation density
$\fs=\ave{\adag_{2'}\av_{1'}}\ket{1'}\otimes\bra{2'}$ and the anomalous
average $\ms=\ave{\av_{2'}\av_{1'}}\ket{1'}\otimes\ket{2'}$ in the
SPE~basis.

Second, we can recognize the fluctuation-density matrix~$\Gs^<$ as the
single-time limit of its position-basis counterpart~$\gs^<(1,2)$ in
Eq.~\eqref{fmsdef}. The mean-field state vector~$\chi$, on the other
hand, can be combined with its Hermitian conjugate into the
matrix~$-i\chi\chi^\dag$, which corresponds to~$h(1,2)$ in
Eq.~\eqref{defh}. We thus define~$\chi(1)\equiv\ave{\hat{A}(1)}$. This
allows us to explicitly connect the condensate mean-field state
vectors~$\chi(1)$ expressed in the position basis and~$\chi(t_1)$ in the
SPE~basis as follows: 
\begin{eqnarray}
  \chi(1)=\left(\begin{array}{cc}\scp{1}{1'}&0\\0&\scp{1'}1\end{array}\right)
	\!\!\left(\begin{array}{c}\al_{1'}(t_1)\\\alpha^\ast_{1'}(t_1)\end{array}\right)
	\equiv T(1)\;\chi(t_1), \label{chitrf1}
\end{eqnarray}
with a time-independent $2\times2n$ transformation
matrix~$T(1)=T(\bx_1)$. Because of the 
completeness of the position basis, we can also write
\begin{equation}
   \chi(t_1)=\int d\bx_1\;T^\dag(1)\chi(1). \label{chitrf2}
\end{equation}
For the fluctuation density, we obtain similarly
\begin{equation}
  i\;\gs^<(1,2)\big|_{t_1=t_2}
   =T(1)\Gs^<(t_1)T^\dag(2) \label{Gstrf1}
\end{equation}
and
\begin{equation}
  -i\;\Gs^<(t_1)=\int d\bx_1\int d\bx_2\;T^\dag(1)\gs^<(1,2)T(2)
	\big|_{t_1=t_2}. \label{Gstrf2}
\end{equation}

We can now use the transformation Eq.~\eqref{Gstrf1} to write the
condensate's Hartree-Fock self-energy~$S\HF (1,\bar1)$ in
Eq.~\eqref{defshf} as
\begin{equation}
  T(1)
     \left(\begin{array}{cc}U_{f^c}+2U_\fs&V_\ms\\
        V_\ms^\dag&U^\dag_{f^c}+2U_\fs^\dag\end{array}\right)
  T^\dag(\bar1)\delta(t_1-t_{\bar1}).\label{Shftrf}
\end{equation}
We here use the definitions of~\cite{Walser2001a}, where energy
shifts due to both the mean field and the normal fluctuations are 
given by the matrices
\begin{equation}
  U_{f}=2\;\phi^{1'2'3'4'}\f_{3'2'}\ket{1'}\otimes\bra{4'},
	\label{Uf}
\end{equation}
whereas the first-order anomalous coupling strength is given by
\begin{equation}
  V_{\ms}=2\;\phi^{1'2'3'4'}\msp_{3'4'}\ket{1'}\otimes\ket{2'}.
	\label{Vm}
\end{equation}
The symmetrized two-body interaction
matrix elements~$\phi$ are here defined by
\begin{equation}
 \phi^{1'2'3'4'}=\frac14 
 (\phi_{u}^{1'2'3'4'}+\phi_{u}^{1'2'4'3'}
 +\phi_{u}^{2'1'3'4'}+\phi_{u}^{2'1'4'3'}),
\end{equation}
\begin{eqnarray}
 \phi_{u}^{1'2'3'4'}=\int d\bx_1\int d&&\bx_2\scp{1'}1\scp{2'}2
\frac{\V{bin}(\bx_1-\bx_2)}{2}\\\nonumber
	&&\times\scp1{3'}\scp2{4'}.
\label{phitrf}
\end{eqnarray}

Like the first-order Hartree-Fock self-energies, we can rewrite the
second-order self-energies in Eqs.\ 
\eqref{defscoll}\ant\eqref{defsigmacoll} using the transformations in
Eqs.\ \eqref{chitrf1}~to~\eqref{Gstrf2}~and~\eqref{phitrf}.
In particular, we now have to transform two potential factors, which
makes the computation more complicated.
Furthermore, the integrals over time to $t_1$\ant$t_2$ in Eqs.\
\eqref{KKM1}~through~\eqref{KKF2} modify one of the binary potentials
according to Eq.~(65) of~\cite{Walser1999a} to an
approximately energy conserving two-particle matrix element
\begin{equation}
  \phi_\eta^{1'2'3'4'}=\phi^{1'2'3'4'}\left\{\pi\delta_\eta(\Delta)
	+i\;{\mathcal{P}}\kern-.2em_\eta\frac1\Delta\right\},
	\label{defphieta}
\end{equation}
with an energy difference~$\Delta$ between the incoming and outgoing
states of the collision event. This definition of the matrix
elements~$\phi_\eta$ introduces the Markov approximation into the
Kadanoff-Baym equations.
We obtain the second-order damping
rates and energy shifts~$\Upsilon^\lgs$ for the condensate,
corresponding to~$S(1,2)$ in Eq.~\eqref{defscoll}, and~$\Gamma^\lgs$
for the fluctuations, corresponding to~$\Sigma(1,2)$ in
Eq.~\eqref{defsigmacoll}; these are the collision integrals defined
in~\cite{Walser2001a}.
The second order terms appear in
combinations~$\Gamma^<\Gs^>-\Gamma^>\Gs^<$ that
contain, for example, the Boltzmann collision terms
\begin{eqnarray}
  \Big\{\Gamma_{\fs\fs\ofs}\ofs&-&\Gamma_{\ofs\ofs\fs}\fs\Big\}_{1'5'}=\\
    8\phi^{1'2'3'4'}\phi_{\eta}^{1''2''3''4''}&\Big\{&
	\fs_{3'1''}\fs_{4'2''}\ofs_{4''2'}\ofs_{3''5'}\nonumber\\
	&-&\ofs_{3'1''}\ofs_{4'2''}\fs_{4''2'}\fs_{3''5'}\Big\}\nonumber
\end{eqnarray}
and similar contributions involving the anomalous averages
$\ms$~and~$\ms^\ast$.

We can now exactly reproduce the coupled equations for the condensed
fraction as well as the normal and anomalous fluctuations stated in Eqs.\
(10)\ant(26) of Ref.~\cite{Walser2001a}. 
Considering the first column of the matrix Eq.~\eqref{KKM1} for the
condensate at $t_1=t_2$, we obtain the generalized Gross-Pitaevskii
equation 
\begin{equation}
  \der{t}\chi=(-i\,\Pi+\Upsilon^<-\Upsilon^>)\chi,	\label{ggpe}
\end{equation}
with the symplectic first-order propagator
\begin{equation}
\Pi=\left(\begin{array}{cc}
	\Pi_{\cal N}&\Pi_{\cal A}\\
	-\Pi_{\cal A}^\ast&-\Pi_{\cal N}^\ast\end{array}\right).
\end{equation}
This propagator consists of the normal Hermitian Hamiltonian 
\begin{equation}
  \Pi_{\cal N}=\Hzero+U_{f^c}+2U_{\fs}-\mu,
\end{equation}
which contains the usual single-particle Hamiltonian~$\Hzero$ given in
Eq.~\eqref{Hzero} and the mean-field and fluctuation shifts~$U_f$
given in Eq.~\eqref{Uf};
furthermore, the symmetric anomalous coupling
\begin{equation}
  \Pi_{\cal A}=V_{\ms}
\end{equation}
is defined in Eq.~\eqref{Vm}.
The propagator~$\Pi$ contains the Hartree-Fock shifts, which are given
in Eq.~\eqref{Shftrf}, and originally were contained in~$S\HF (1,2)$
[see Eq.~\eqref{defshf}].

To obtain the equation of motion for the
fluctuations, we subtract Eq.\ \eqref{KKF1}~from~\eqref{KKF2} and
evaluate at $t_1=t_2$ to obtain 
\begin{equation}
  \der{t}\Gs^<=-i\,\Sigma\Gs^<+\Gamma^<\Gs^>-\Gamma^>\Gs^<+\hc
	\label{gbe}
\end{equation}
The reversible evolution of the fluctuations~$\Gs^<$ 
is governed by the Hartree-Fock-Bogoliubov self-energy operator
\begin{equation}
  \Sigma=\left(\begin{array}{cc}\Sigma_{\cal N}&\Sigma_{\cal A}\\
	-\Sigma_{\cal A}^\ast&-\Sigma_{\cal N}^\ast\end{array}\right),
\end{equation}
which in turn consists of the Hermitian Hamiltonian
\begin{equation}
  \Sigma_{\cal N}=\Hzero+2U_{f^c}+2U_{\fs}-\mu
\end{equation}
and the symmetric anomalous coupling
\begin{equation}
  \Sigma_{\cal A}=V_{m}.
\end{equation}
The propagator~$\Sigma$ corresponds to~$\Sigma\HF (1,2)$ in
Eq.~\eqref{defsigmahf}. Its mean-field shift is twice as
large as that of the condensate propagator~$\Pi$, which is a well
known property of first-order Hartree-Fock-Bogoliubov theories.
Further details of this transformation can be found
in~\cite{Wachter2000a}.

\section{Conclusion}
We independently rederive the kinetic equations of Walser {\it et al.}\ from
the Kadanoff-Baym nonequilibrium Green's function formulation of
kinetic theory, and recover identical factors in all second-order
damping rates and energy shifts. This shows that for dilute, weakly
interacting gases the Kadanoff-Baym nonequilibrium, real-time Green's
function approach is microscopically equivalent to the density matrix
approach used by Walser {\it et al.}~\cite{Walser2001a}. The latter approach
is more physical in two respects: First, its variables are measurable
quantities: the mean field and the density and coherences of thermal
atoms. Second, the variables' equations of motion reduce to the
Gross-Pitaevskii equation and the quantum Boltzmann equation in the
low- and high-temperature limits, respectively.

Starting from the gapless Beliaev approximation for the collisional
self-energy in the Kadanoff-Baym equations, we furthermore learn that
the full second-order kinetic theory of Walser~{\it et al.}\ is gapless
itself~\cite{Beliaev1958a,Imamovic2000a}.  This shows that the gap
that appears in the first-order Hartree-Fock-Bogoliubov
spectrum~\cite{Griffin1996a} is closed by the second-order energy
shifts.

Furthermore, this work connects the kinetic theory of Walser {\it et al.}\
with work done by M. Imamovi\'c-Tomasovi\'c {\it et
al.}~\cite{Imamovic2000a,Imamovic2000b,Imamovic1999a}, because they 
start from the same Kadanoff-Baym equations.

\section*{Acknowledgments}
J.W. acknowledges financial support by the National Science
Foundation. 
R.W. gratefully acknowledges support from the Austrian
Academy of Sciences through an APART grant.
M.H. acknowledges support from the U.S. Department of Energy, Office
of Basic Energy Sciences via the Chemical Sciences, Geosciences, and
Biosciences Division.


\ifpreprintsty\else\end{multicols}\fi
\end{document}